\documentclass[preprint]{iucr}              
\usepackage{amsmath}

     \journalcode{J}              

\begin{document}                  




\title{Optimal weights and priors in simultaneous fitting of multiple small-angle scattering datasets}

\cauthor[a]{Andreas Haahr}{Larsen}{andreas.larsen@sund.ku.dk}{address if different from \aff}
\aff[a]{University of Copenhagen, Department of Neuroscience, Blegdamsvej 3, 2200 Copenhagen, \country{Denmark}}

\maketitle

\begin{synopsis}
Determining optimal weights and priors when simultaneously fitting SAXS and SANS data.
\end{synopsis}

\begin{abstract}
Small-angle X-ray and neutron scattering (SAXS and SANS) are powerful techniques in material science and soft matter. In this study, it was addressed how multiple SAXS or SANS datasets are best weighted when doing simultaneous fitting. Three weighting schemes were tested: (1) equal weighting of all datapoints, (2) equal weighting of each dataset through normalization with the number of datapoints, (3) weighting proportional to the information content. The weighing schemes were assessed by model refinement against synthetic data under numerous conditions. The first weighting scheme led to the most accurate parameter estimation, especially when one dataset substantially outnumbered the other(s). Furthermore, it was demonstrated that inclusion of Gaussian priors significantly improved the accuracy of the refined parameters, as compared to common practice,  where each parameter is constrained uniformly within an allowed interval. 

\end{abstract}

\section{Introduction}
Small-angle X-ray and neutron scattering (SAXS and SANS) provide structural information about nanoscale structures, ranging from a few to hundreds of nanometers. It has applications across diverse fields, including investigations of  amorphous materials like gels, polymers and glasses, as well as biological macromolecules such as proteins, DNA, lipids and their complexes. Hard materials, including nanoparticles also fall within the scope of investigation. By combining SAXS or SANS measurements which have different scattering length contrasts, structural domains can be highlighted, resulting in more accurate refinement of structural parameters.

Contrast variation can be achieved in SAXS by changing the ionic strength of the solvent  \cite{Gabel2019}, or by ASAXS \cite{Ballauff2006}. In SANS, the contrast can be varied using hydrogen-deuterium exchange in sample or solvent \cite{Heller2010}. SAXS and SANS have elegantly been combined, e.g., in studies of toroidal polymers assemblies \cite{Hollamby2016}, protein/DNA complexes \cite{Sonntag2017}, multishell-nanoparticles \cite{Lin2020}, block copolymer micelles \cite{Manet2011}, multilamellar lipid vesicles \cite{Heftberger2013}, and  lipid nanodiscs \cite{Kynde2014-bR}, to mention a few examples. 
However, choosing proper weights to each dataset is not trivial: should one simply weight with the number of points, or should the number of points be normalized out in the minimization? Should the noise level and information content be taken into account in the minimization algorithm? In this paper, three weighting schemes were compared: (1) a naive weighting scheme, where each datapoint is weighted according to its statistical uncertainty with no additional weighting, meaning that datasets with more points have more weight; (2) a so-called reduced weighting scheme, where each dataset is given equal weight, corresponding to minimizing the reduced $\chi^2$, and (3) an information-based weighting scheme, where each dataset is weighted proportional to their information content. Model parameters were co-refined against synthetic data, and the refined values were compared to the known ground truth to evaluate and compare the different weighting schemes. 

Another central aspect in modeling, is the inclusion of molecular constraints \cite{Zemb2010} or prior knowledge.
The present study advocates for the use of Bayesian refinement with Gaussian priors for enhanced accuracy in co-refinement against multiple SAXS or SANS datasets. This is inspired by successful applications  of Bayesian refinement in X-ray crystallography  \cite{Headd2012-Phenix}, electron microscopy \cite{Scheres2012-Bayes,Scheres2012-RELION}, reflectometry \cite{Nelson2019,McCluskey2020,McCluskey2023}, and for the combining of SAXS with molecular dynamics simulations \cite{Hummer2015}.

\section{Methods}
This paper relies on fitting simulated or synthetic data. Thus, the ground truth is known, allowing for quantitative evaluation of different weighting schemes and prior inclusion. For the generation and analysis of synthetic data, two form factors were applied.

\subsection{Core-multishell form factor}
The core-shell model is built up using the form factor amplitude for a sphere with radius $R$:
\begin{align}
	\psi_s(qR) = 3 \frac{\sin(qR)-qR\cos(qR)}{(qR)^3}
\end{align}
The amplitude of the scattering vector is $q=4\pi\sin(\theta)/\lambda$, where $2\theta$ is the scattering angle and $\lambda$ is the wavelength of the incoming wave. The volume of the sphere is $V_s(R)=4\pi R^3/3$. The core radius of the model is denoted $R_\mathrm{c}$, and the outer radius of the $i^{th}$ shell is denoted $R_i$. The difference in scattering length density between the $i^{th}$ shells and the solvent, i.e. the contrast, is denoted $\Delta\rho_i$. The form factor for a core-multishell particle with $n_s$ shells can be written as:
\begin{align}
	P_\mathrm{cs}(q) = \left[\frac{V_\mathrm{s}(R_\mathrm{c})\psi_s(q,R_\mathrm{c}) + \sum_{j=1}^{n_s}\frac{\Delta\rho_j}{\Delta\rho_\mathrm{c}}\left[V_\mathrm{s}(R_j)\psi_\mathrm{s}(qR_j) - V_\mathrm{c}(R_{j-1})\psi_\mathrm{s}(qR_{j-1})\right]}{V_\mathrm{s}(R_\mathrm{c}) +  \sum_{j=1}^{n_s}\frac{\Delta\rho_j}{\Delta\rho_\mathrm{c}}(V_\mathrm{s}(R_j) - V_\mathrm{s}(R_{j-1})) }\right]^2
	\label{eqn:P_cms}
\end{align}
For this paper, we used three shells ($n_\mathrm{s}=3$). The intensity is modeled with a scaling and a constant background, $I_\mathrm{cs}(q)=a P_\mathrm{cs}(q)+b$. Only the \emph{relative} values of the contrasts affect $P(q)$, so the model has 9 parameters ($K=9$): four radii ($R_\mathrm{c},R_1,R_2,R_3$), three relative contrasts ($\Delta\rho_i/\Delta\rho_c)$,  as well as scaling and constant background. When fitting two datasets with the model, five additional parameters were introduced, namely three relative contrasts, scaling, and background for the second dataset ($K=14$). 

\subsection{Stacked cylinder form factor}
For testing the method against a less symmetric model with a different contrast situation, a stacked cylinder form factor was used. The model is based on the form factor amplitude for cylinders with radius $R$ and length $L$ \cite{Pedersen1997}: 
\begin{align}
		\psi_c(q,R,L,\alpha) = \frac{2B_1(qR\sin\alpha)}{qR\sin\alpha}\frac{\sin(qL\cos\alpha/2)}{qL\cos\alpha/2}
\end{align}
This form factor amplitude should be integrated over $\alpha$ to yield the cylinder form factor. The volume of the cylinder is $V_\mathrm{c}(R,L)=\pi R^2 L$. The form factor for $n_c$ stacked cylinders with radii $R_i$, lengths $L_i$ and contrasts $\Delta \rho_i$ is:
\begin{align}
	P_{c}(q) = \int_0^{\pi/2} \left|\frac{ \sum_{j=1}^{n_c} \frac{\Delta\rho_j}{\Delta\rho_1} V_\mathrm{c}(R_j,L_j) \psi_c(q,R_j,L_j,\alpha) \phi_j(\alpha,L_1,...,L_j)}{\sum_{j=1}^{n_c} \frac{\Delta\rho_j}{\Delta\rho_1} V_\mathrm{c}(R_j,L_j)}\right|^2 \sin\alpha d\alpha
\end{align}
where $\phi_i$ is the phase factor of the $j^{th}$ cylinder, which depends on the center-to-center distance to the first cylinder:
\begin{align}
	\phi_j(\alpha,L_1,...,L_j) = \exp\left(iq \left(-\frac{L_1+L_j}{2} + \sum_{k=1}^{j} L_k\right) \cos\alpha\right)
\end{align}
In the special case $j=1$ then $\phi_j$ is unity. For this paper, we used three stacked cylinders ($n_c=3$), each with the same radii, but with varying lengths. The intensity was modeled with a scale and a background, $I_\mathrm{c}(q)=a P_c(q)+b$. This model had 7 parameters ($K=7$), when refined against a single dataset, and eleven parameters ($K=11$), when two datasets were simultaneously fitted. 

\subsection{Model implementation and validation}
The form factors were implemented in BayesFit (github.com/andreashlarsen/BayesFit), and validated against simulated data generated in Shape2SAS \cite{Shape2SAS}.

\subsection{Simulated SAXS and SANS data}
First, $q$ was defined, with $q_\mathrm{min}=0.001$ \AA$^{-1}$ and $q_\mathrm{max}=0.5$ \AA$^{-1}$ for the spherical core-multishell particles and $q_\mathrm{min}=0.0001$ \AA$^{-1}$ and $q_\mathrm{max}=0.3$ \AA$^{-1}$ for the stacked cylinders. The simulated SANS-like data contained 50 or 300 points, and the simulated SAXS-like data contained either 300, 400, 900 or 2000 points. Theoretical curves were then calculated and evaluated at these $q$-values, using $I_\mathrm{model}(q)=a P(q)+b$. The SAXS data were scaled by $a_\mathrm{SAXS}=0.5$ cm$^{-1}$ and the SANS data by $a_\mathrm{SANS}=0.8$ cm$^{-1}$, and a constant background of $b=10^{-5}$ cm$^{-1}$ was added to the SAXS data and $b=10^{-4}$ cm$^{-1}$ was added to the SANS data. To ensure realistic errors, similar to what would be obtained from an experiment, the errors were modelled using an empirical model \cite{Sedlak2017-errors}: 
\begin{align}
	\sigma_i = \sqrt{\frac{I_s(q_i) + 2cI_s(0.2 \mathrm{\AA}^{-1})/(1-c)}{4500q_i}},
\end{align}
where $I_s(q_i) = sI_\mathrm{model}(q_i)/I(0)$ is the normalized and scaled model intensity evaluated at $q_i$, and $\sigma_i$ are the standard deviations, which in an experiment are estimated through counting statistics and error propagation. For simulated SAXS-like data, $s=100$ and $c=0.85$ were used (high signal-to-noise ratio), whereas for simulated SANS-like data, $s=10$ and $c=0.95$ were used (lower signal-to-noise ratio). The simulated intensities ($I_i$) were then pulled stochastically from normal distributions with mean $\mu_i=I_\mathrm{model}(q_i)$ and standard deviation $\sigma_i$.  

To simulate data with increasing noise, the variance ($\sigma_i^2$) was multiplied with a noise factor, before simulation of the intensities, i.e. $\sigma_i^2 \rightarrow f_\mathrm{noise}\sigma_i^2$. The noise was increased logarithmically, by varying $\log(f_\mathrm{noise})$ from $-4$ to $10$. In order to simulate data with over- or underestimated errors, a factor was multiplied on $\sigma_i$ after simulation of the data, such that $\sigma_i$ no longer reflected the fluctuations of the simulated intensities \cite{Smales2021,Larsen2021-chi2}. For each condition, i.e., the different weight schemes and priors described in the results section, 50,000 SAXS and 50,000 SANS datasets were simulated and fitted with the model. 

Due to wavelength spread, divergence and pixel size, there are instrumental smearing effects or resolution effects \cite{Pedersen1990}. These are usually negligible in synchrotron SAXS data, but not in SANS and lab-source SAXS data. Based on the instrumental settings, the resolution effects can in many cases be expressed as a normal distributed error, $\sigma_q$ for each $q$-value, and included in the model by smearing the theoretical intensity:
\begin{align}
	I_\mathrm{model, res}(q) =  \frac{1}{\sigma_q\sqrt{2\pi}} \int_{-\infty}^{\infty} I_\mathrm{model}(q') \exp\left(-\frac{1}{2}\left(\frac{q'-q}{\sigma_q}\right)^2\right) dq'
\end{align}
At many SANS instruments, the values of $\sigma_q$ are provided as a fourth column in the datafile. To investigate the effect of smearing, the fourth column of a SANS dataset from D22 was used (SASBDB: SASDL53) \cite{Lycksell2021}. The experimental $\sigma_q$ values were imported and linearly extrapolated to the simulated $q$ values. To investigate the effect of larger resolution effects, data were also simulated with $\sigma_q$ multiplied by a factor of 2 or 3. The resolution effects were taken into account when fitting these data, using the same $\sigma_q$ values that were used to simulated data.

\subsection{BayesFit - fitting multiple datasets with priors}
BayesFit (github.com/andreashlarsen/BayesFit) is a program that can fit SAXS and SANS data simultaneously with an analytical model, and use Gaussian priors. BayesFit was originally implemented in FORTRAN \cite{Larsen2018-BayesFit}. For this paper, a new implementation was written in Python, to facilitate fitting of multiple datasets. BayesFit reads an input file, which contains information about the data, the name of the model, the prior values for each model parameter ($\mu_{\mathrm{prior},k}$ and $\sigma_{\mathrm{prior},k}$) and the weights ($w_j$) used to balance different datasets. The weight given to the prior is adjusted by a hyperparameter, $\alpha$ \cite{Hansen2000,Larsen2018-BayesFit}. BayesFit minimizes:
\begin{align}
	\min\left[\left(\sum_{j=1}^{N_\mathrm{dataset}}w_j\chi^2_j\right) + \alpha S\right],
	\label{eqn:BayesFit}
\end{align}
where $\chi^2$ and $S$ are given as:
\begin{align}
	\chi^2 &=\sum_{i=1}^{M} \left(\frac{I_i-I_\mathrm{model}(q_i)}{\sigma_i}\right)^2,\label{eqn:chi2}\\
	S         &=\sum_{k=1}^{K} \left(\frac{x_k-\mu_{\mathrm{prior},k}}{\sigma_{\mathrm{prior},k}}\right)^2.
	\label{eqn:S}
\end{align}
$x_k$ is the refined value of the $k^{th}$ model parameter and $M$ is the number of datapoints. For the refinements in this paper, BayesFit scanned 11 logarithmically spaced values of $\alpha$ and the range was manually adjusted to ensure that it contained the $\alpha$ values giving rise to the highest probabilities. BayesFit utilizes Scipy's curve\_fit function \cite{Virtanen2020-scipy}. In order to use the curve\_fit function, an array was defined with all $q$-values from both SAXS and SANS data, and dummy $q$-values for each of the prior values. A corresponding array was defined with all simulated intensities ($I_i$) from the SAXS and SANS datasets and the prior means ($\mu_{\mathrm{prior},k}$). An array was finally constructed, with the errors of the simulated data ($\sigma_i$) as well as the prior standard deviations ($\sigma_{\mathrm{prior},k}$). The experimental errors were scaled with $w_j^{-1/2}$ before fitting, to obtain the weighting in equation \eqref{eqn:BayesFit}. The prior means ($\mu_{\mathrm{prior},k}$) were used as initial guesses in the subsequent nonlinear minimization. The upper and lower limits were set to $\pm5\sigma_{\mathrm{prior},k}$, and parameters were constrained to positive values when relevant. To apply uniform priors, $\alpha$ was fixed at $10^{-10}$, effectively quenching the effect of the prior, except for the upper and lower limits, which were adjusted by changing $\sigma_{\mathrm{prior},k}$. The means, $\mu_{\mathrm{prior},k}$, were also used as initial guesses when fitting with uniform priors. Parameter values for all priors are listed in Table \ref{table:R} and Table \ref{table:p}. Normalized Hessian matrices and their eigenvalues were used to calculate the information content \cite{Vestergaard2006}. The Hessian matrices were constructed numerically from $\chi^2$ using the forward Euler method,  and eigenvalues were found using NumPy \cite{Harris2020-numpy}. The total probability of the solution, taking into account the likelihood and priors, were derived from Bayes theorem \cite{Hansen2000,Larsen2018-BayesFit}. Each refined model parameter was then calculated as a probability-weighted average:
\begin{align}
	x_{\mathrm{refined},k} = \sum_{i=1}^{N_\alpha}p(\alpha_i)x_k(\alpha_i),
\end{align}
where $p(\alpha_i)$ is the probability density of the solution at $\alpha_i$, and $x_k(\alpha_i)$ is the refined value of the $k^{th}$ parameter at $\alpha_i$. $N_\alpha$ is the number of $\alpha$ values that were scanned. 
The program is meant as a proof-of-concept, and the goal is that inclusion of Gaussian priors and optimal weighting should be implemented in other software packages for SAXS and SANS analysis, which are superior in the number of verified models, user interface, performance and additional features. Such programs include WillItFit \cite{Pedersen2013-WillItFit} and SasView (www.sasview.org). From SasView version  6, it was made possible to adjust weights in simultaneous fitting (www.sasview.org/downloads/modifying\_weights\_in\_sasview\_v6.pdf), which calls for thorough investigations of which weighting scheme is most optimal.

\subsection{Calculating information content}
The number of good parameters ($N_\mathrm{g,BIFT}$) was used as a measure for the information content in data. $N_\mathrm{g,BIFT}$ was chosen instead of the number Shannon channels \cite{Shannon1949,Nyquist1928}, as $N_\mathrm{g,BIFT}$ takes into account the noise level of data \cite{Vestergaard2006} (Figure S1). $N_\mathrm{g,BIFT}$ was calculated with a Bayesian indirect Fourier transformation (BIFT) algorithm \cite{Hansen2000}, as implemented in BayesApp (version1.1) \cite{Hansen2012,Larsen2021-chi2}. BIFT cannot fit all data, so one may in those cases replace $N_\mathrm{g,BIFT}$ by the number of Shannon channels. 

\subsection{Estimating degrees of freedom to calculate reduced $\chi^2$  values}
The number of good parameters is a good measure for the degrees of freedom ($DOF$) in a fit and therefore provide a correct estimate of the reduced $\chi^2$, namely $DOF = M-N_g$, where $M$ is the number of datapoints \cite{Larsen2018-BayesFit,Larsen2021-chi2}. That is also the case for simultaneous fitting against multiple data (Figure S2). However, it is not evident what the degrees of freedom (and reduced $\chi^2$ values) should be for each dataset in a simultaneous fit. The number of good parameters for each dataset ($N_{g,j}$) should sum up to the total $N_g$ for the simultaneous fit. An upper limit of $N_{g,j}$ can be  estimated following the usual approach \cite{Larsen2018-BayesFit} for each dataset, and is denoted $n_{g,j}$. By requiring that the sum of $N_{g,j}$ values should equal the total $N_g$, we reach that
\begin{align}
	N_{g,j} = n_{g,j} - \frac{\Sigma  n_{g,j} -  n_{g,j} }{\Sigma  n_{g,j} }(\Sigma  n_{g,j} - N_g),
\end{align}
This is a good measure for the degrees of freedom, as assessed by monitoring the reduced $\chi^2$ from simultaneously fitting against simulated data (Figure S3).

\subsection{Molecular dynamics simulations}
The deposited structure "model-1 (pdb)" (SASBDB entry SASDNK2) was used as initial frame. The structure was solvated in TIP3P water with 100 mM NaCl, in a cubic box with box lengths 27 nm and periodic boundary conditions. Simulations were run in GROMACS 2021.4 with force fields AMBER14SB\_OL15 or CHARMM36-IDP. The structure was minimized, then equilibrated with constant number of particles, volume and temperature (NVT) for 100 ps, then with constant number of particles, pressure and temperature (NPT) for another 100 ps. The protein was position restraint during these equilibration steps. Temperature 300 K, time constant 0.1 ps kept with the v-rescale algorithm. Pressure was kept at 1 bar using Parrinello-Rahman pressure coupling and a time constant of 2 ps.  The restraints were released and the simulation was run for 100 ns with NPT. 

\subsection{Calculating theoretical scattering from the molecular dynamics simulations}
The first 40 ns of the simulations were excluded to avoid the results being dependent on the initial frame. The theoretical scattering was calculated from the remaining 60 ns with Pepsi-SANS (for Linux) version 3.0 (https://team.inria.fr/nano-d/software/pepsi-sans). For the SANS data, the scattering from the KaiA domain only was compared to data, as the KaiB and KaiC domains were matched out in the experiment. 

\section{Results}
The results section contains two parts. In the first, it is investigated which weighting schemes is best, when simultaneously fitting multiple SAXS or SANS contrasts. In the second part, the inclusion of priors is investigated.

\subsection{Finding the best weighting scheme}
When refining a model against multiple datasets, e.g. a SAXS and a SANS dataset, or multiple SANS contrasts, a central question is how to weight each datasets. The model refinement is done by minimizing the weighted sum:
\begin{align}
	\min\left[\sum_{j=1}^{N_\mathrm{dataset}}w_j\chi^2_j\right],
	\label{eqn:weight}
\end{align}
where $\chi^2$ is defined in equation \eqref{eqn:chi2}. Assuming independent datapoints, the sum of $\chi^2$ should be minimized with no additional weighting, i.e., $w_j=1$. This naive weighting scheme is the first that will be tested. However, equation \eqref{eqn:weight} is a sum over the non-reduced $\chi^2$, which scales with the number of datapoints, so the result is dominated by the larger dataset. To counteract this, one may use the weight, $w_j=1/M_j$. This is the second weighting scheme that will be tested, and roughly corresponds to replacing $\chi^2$ with the reduced $\chi^2$ in equation \eqref{eqn:weight}, so it will be denoted the reduced weighting scheme. A third approach is to weight by the information content in data, e.g., by the number of good parameters $N_\mathrm{g,BIFT}$ \cite{Vestergaard2006}. That way, the data with the highest information content also get the highest weight, i.e., $w_j=N_{\mathrm{g,BIFT},j}/M_j$. A similar information-based weighting scheme has previously been applied to combine SAXS and molecular dynamics simulations \cite{Shevchuk2017}. 

\begin{figure}
	\caption{Refinement of a core-multishell model using different weighting schemes. (A) Core-multishell particle with relative contrasts and radii annotated. (B) Simulated SAXS-like data with 400 points. (C) Simulated SANS-like data with 50 points. (D-G) Refined values of $R_\mathrm{c}$, $R_1$, $R_2$ and $R_3$ from 50,000 fits (new data simulated each time). The parameters were either refined against SANS alone (green area), SAXS alone (red area), or SAXS and SANS with naive weighting scheme (red line), reduced weighting scheme (green line) or information-based weighting scheme (red line). The gray vertical line is the true value.}
	\label{fig:R_w}
	\includegraphics[width=0.90\linewidth]{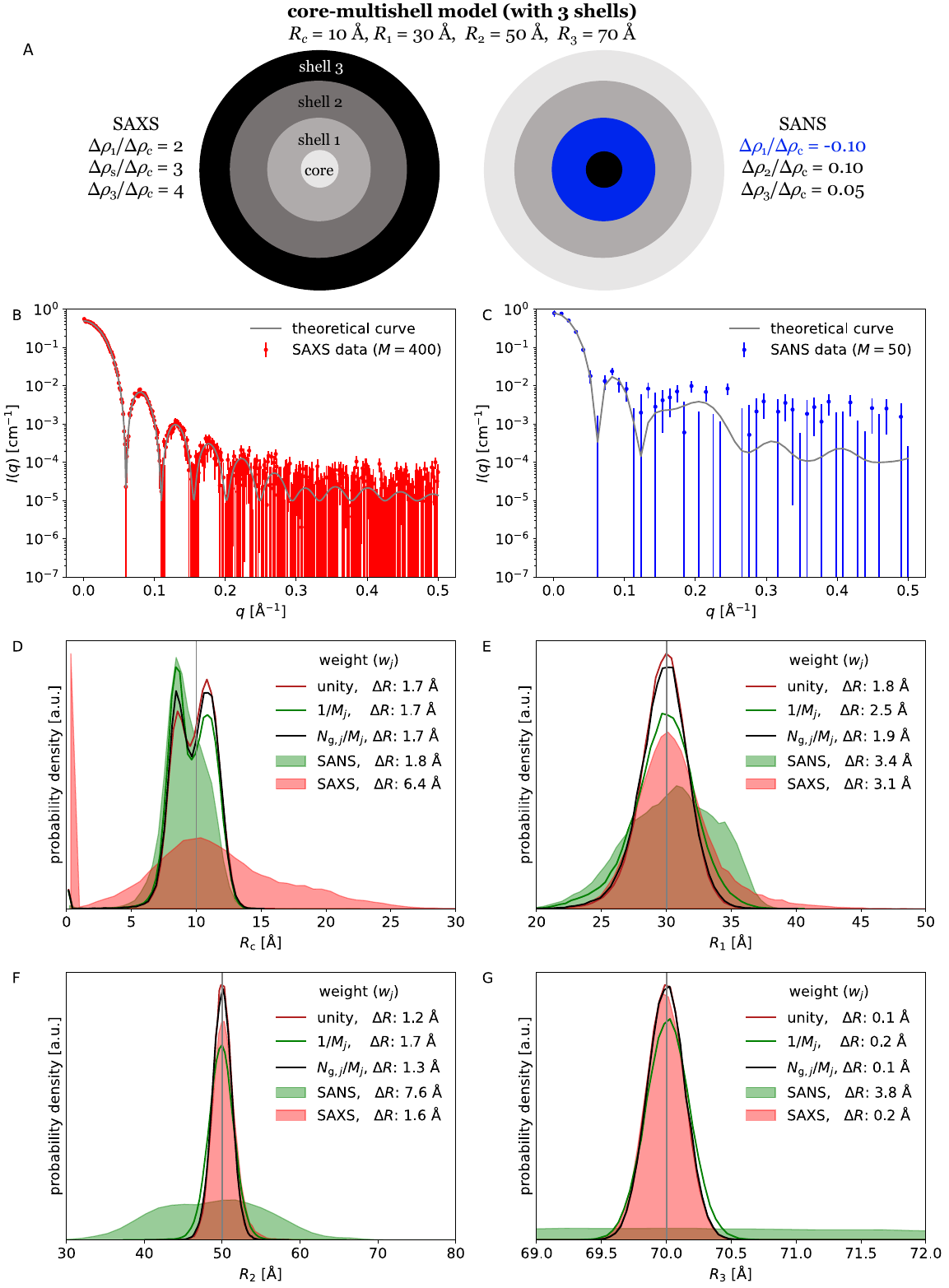}
\end{figure}

In order to test which weighting scheme performs best, two datasets were simulated for a sample of core-multishell particles. The particles had three shells, so a total of four radii were refined from the data. The true values were 10, 30, 50 and 70 Å. The first dataset contained 400 datapoints with a relatively high signal-to-noise ratio, and the second dataset contained only 50 datapoints and a lower signal-to-noise ratio. These data mimic an experiment, where the sample is measured with two different contrast situation, e.g. with synchrotron SAXS and with SANS (Figure \ref{fig:R_w}). Most SANS data contain more points than 50 and often there will be multiple SANS contrast, so the total amount of SANS datapoints could often exceed the number of SAXS datapoints. However, the low number was chosen to explore a situation with substantial difference between the size of the two datasets, i.e., where the weight schemes are more important. The true model that was used to generate the simulated data, were then refined against the simulated SAXS-like and SANS-like datasets using the three weighting schemes $w_j=1$, $w_j=1/M_j$, or $w_j=N_{\mathrm{g,BIFT},j}/M_j$, to estimate the geometric parameters and compare with the true values. The model parameters were also refined against  SAXS data alone and SANS data alone. To mimic an experiment, the simulated data were generated stochastically. Therefore, the simulation and analysis protocol was repeated 50,000 times ($n_\mathrm{rep}$) for each weighting scheme, to get a distribution of refined parameter values. The best weighting scheme is the one that gives the most accurate parameters values after refinement, i.e., closest to the ground truth. To quantify the accuracy of the determination of each parameter, the deviation from the true value was defined as: 
\begin{align}
	\Delta x_j = \sqrt{\frac{1}{n_\mathrm{rep}}\sum_{i=1}^{n_\mathrm{rep}}(x_\mathrm{j,true}-x_{\mathrm{j,refined},i})^2}.
\end{align}
Since the true value is known, there are zero degrees of freedom, and the denominator is $n_\mathrm{rep}$ and not $n_\mathrm{rep}-1$ as in the standard deviation, where the true value must be estimated as the mean. We use the relative deviations $\Delta x_j/|x_{j,\mathrm{true}}|$, to calculate an average relative deviation of a set of parameters:
\begin{align}
	 \text{average\ relative\ deviation} = \frac{1}{K} \sum_{j}^{K} (\Delta x_j/|x_{j,\mathrm{true}}|)
	 \label{eqn:av_dev}
\end{align}

\paragraph{Which weighting scheme is best for refinement of the core-multishell model?}
This can be answered by comparing how accurately the structural parameters of the core-multishell model were refined with the different weighting schemes. The radius of the core ($R_\mathrm{c}$) was ill-determined by the data due to the contrast situation (Figure \ref{fig:R_w}A). Therefore, it was not uniquely determined using any of the weighting schemes (Figure \ref{fig:R_w}D). The average relative deviation from the true value, $\Delta R_c$, was 1.7 \AA\ irrespective of applied weighting scheme, so no weighting scheme was substantially better than the others for this parameter. However, the outer radius of the first and second shells ($R_1$ and $R_2$) were refined most accurately when using the naive weighting scheme $w_j=1$ (simply using experimental errors as weights), closely followed by the information-based weighting scheme $w_j=N_{\mathrm{g,BIFT},j}/M_j$ (weighting with information content), whereas when using the reduced weighting scheme $w_j=1/M_j$ (corresponding to using reduced $\chi^2$ instead of $\chi^2$), the refined values were substantially less accurate (Figure \ref{fig:R_w}E-F). For the outer radius of the third shell ($R_3$), the naive weighting scheme ($w_j=1$) and the information-based weighting scheme ($w_j=N_{\mathrm{g,BIFT},j}/M_j$) resulted in equally accurate results (Figure \ref{fig:R_w}G). 

In order to assess the accuracy of a given weighting scheme using a single number, the average deviation across the radii were calculated, as in equation $\eqref{eqn:av_dev}$. The average deviation across all radii  were 6.4\% for the naive weighting scheme ($w_j=1$), 6.5\% for the information-based weighting scheme ($w_j = N_{g,j}/M_j$) and 7.8\% for the reduced weighting scheme ($w_j=1/M_j$). So the naive weighting scheme performed best for these data, as the average deviation was smallest.

\begin{figure}
	\caption{Refinement of a stacked cylinder model against simulated data, using different weighting schemes. (A) Stacked cylinders with dimensions and relative contrasts annotated. (B) Simulated SAXS-like data with 400 points. (C) Simulated SANS-like data with 50 points. (D-G) Histograms of refined values of $R$, $L_1$, $L_2$ and $L_3$ (gray line is the true value), after simultaneous fits to 50,000 pairs of simulated SAXS and SANS data.}
	\label{fig:cyl}
	\includegraphics[width=0.95\linewidth]{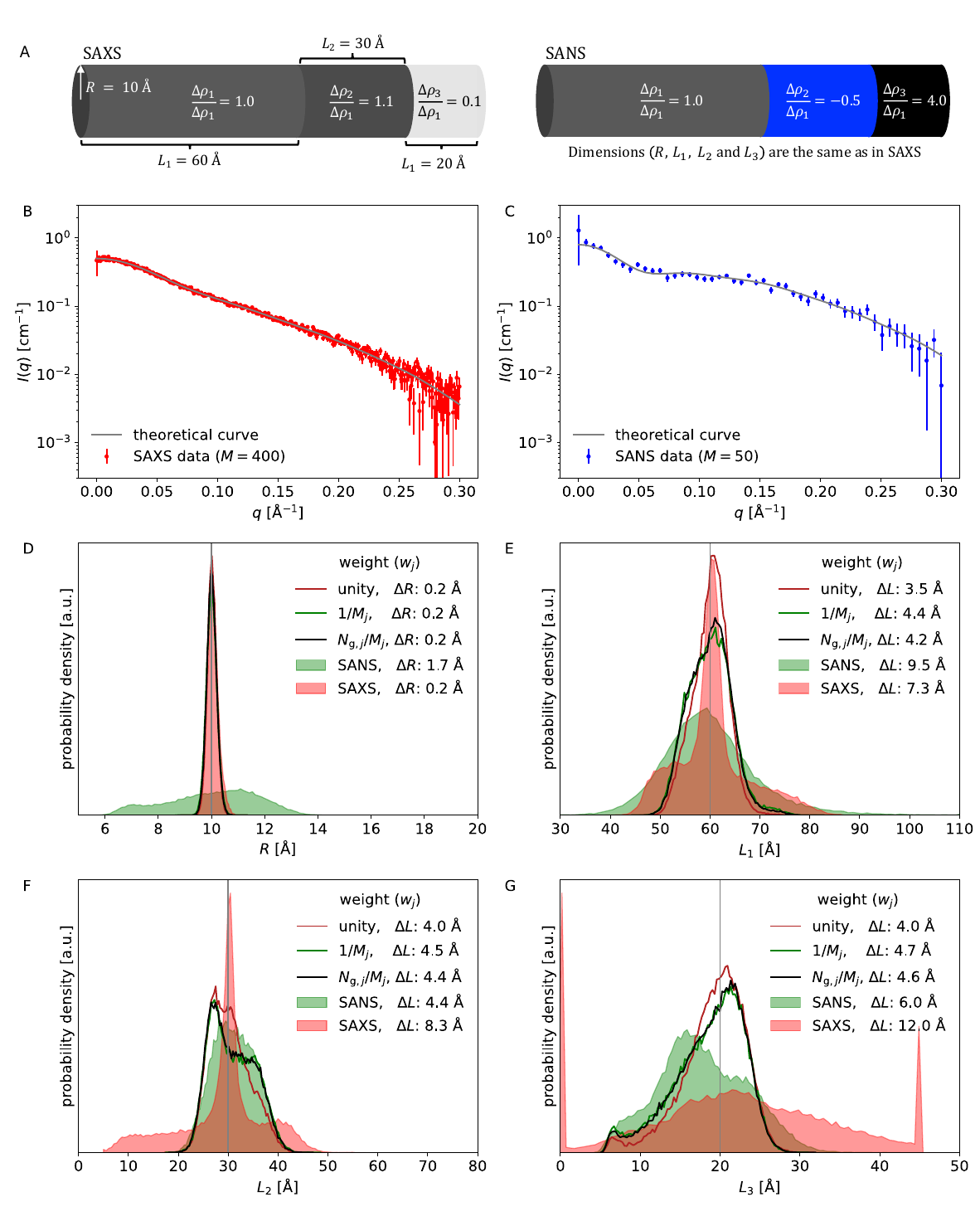}
\end{figure}

To investigate the generality of the result, other conditions were tested using the same approach, as summarized in Table \ref{table:DR}. This included changing the number of points in each dataset, adding a SANS dataset for highlighting the core radius, and adding interparticle interactions. The effect of an inaccurate model as well as resolution effects were also investigated. This were all done with the spherical core-multishell model (Figure \ref{fig:R_w}). Finally, the weighting schemes were evaluated against a stacked cylinder model (Figure \ref{fig:cyl}). 

\paragraph{Effect of changing the number of points in each contrast. }
To investigate the effect of the number of points in data, the same spherical core-multishell model was used, but new pairs of SAXS- and SANS-like data were simulated with the number of points in the datasets being varied. The ratios of points in the two datasets spanned from 1:1 (300 points in each dataset) to 1:40 (respectively 50 and 2000 points). When the number of points were the same, all weighting schemes performed equally well. However, as the difference in number of points increased, the naive weighting scheme ($w_j=1$) gave the most accurate results (Table \ref{table:DR}). Intriguingly, all weighting schemes were superior to fitting against SAXS or SANS data alone. The ratio between number of points in each dataset had to be high (a factor of 6 or higher), before there were substantial difference between the naive weighting scheme ($w_j=1$) and the information-based weighting scheme ($w_j = N_{g,j}/M_j$), whereas the reduced weighting scheme ($w_j = 1/M_j$) always resulted in less accurate parameter refinement (Table \ref{table:DR}, rows 1-4).

\paragraph{More than two contrasts included. }
Additional contrasts are often measured if the sample contains multiple internal scattering length densities. Therefore, a SANS-like contrast was simulated, where only the core had non-zero contrast with respect to the buffer. The spherical core-multishell model was then fitted against the two original datasets (Figure \ref{fig:R_w}) and the new SANS core contrast. Unsurprisingly, this addition dramatically improved the accuracy of the core radius refinement, $R_c$ (Figure S4). However, the conclusions regarding the choice of weighting scheme remained the same; the naive weighting scheme ($w_j=1$) gave the most accurate refinement, especially when there were significant difference between the number of datapoints in each contrast (Table \ref{table:DR}, rows 5-6). 

\paragraph{Interparticle interactions. }
If there is interparticle interactions and correlation between the location of individual particles, a simple form factor is not a sufficient description, and addition of a structure factor is necessary. To investigate that situation, data were simulated with a hard-sphere structure factor to consider interparticle interactions of highly concentrated samples. The same hard-sphere structure factor was used when fitting the data. For the combination of a simulated SAXS dataset with 400 points and a simulated SANS dataset with 50 points, the information-based weighting scheme ($w_j = N_{g,j}/M_j$) had the smallest deviation from the true parameter values. However, as the difference in number of points between the datasets increased, the naive weighting scheme ($w_j=1$) gave the smallest average deviation (Table \ref{table:DR}, rows 7-8). 

\paragraph{Systematic errors: inaccurate models and resolution effects. }
Examples of systematic errors include interparticle interactions where the structure factor is assumed to be unity, aggregation or oligomerization of a sample that is assumed to be monodisperse, or roughness of surfaces that are modeled as smooth. Systematic errors may also stem from undesired experimental effects, including reflections from the sample holder or buffer mismatches. 

To investigate one of these systematic errors, data were simulated using a model with a raspberry-like surface. This model was similar to the core-multishell model, except that the outer shell (shell number 3) was removed and instead, the surface of shell number 2 was covered by small spheres. The data were, however, still fitted with the simpler core-multishell model. So the data were simulated with one model, but fitted with a simpler, inaccurate model. This resulted in large variation of the refined values (Table \ref{table:DR}, rows 9-10) due to ambiguous determinations of the outer two shells (Figure S5). However, despite the inaccurate model, the naive weighting scheme ($w_j=1$) remained the most accurate (Table \ref{table:DR}, rows 9-10).

Resolution effects is another important aspect to consider, especially in SANS. As neighboring points are related through smearing effects, one may suspect that the naive weighting scheme ($w_j=1$), which assumes independent datapoints, would perform worse. Therefore, resolution effects were applied to the simulated SANS data. The resolution effects were likewise included in the subsequent fitting process. The resolution effects, which are described as an uncertainty in $q$, were multiplied by factors of 2 or 3 to simulate more severe resolution effects. In all cases, however, the naive weighting scheme ($w_j=1$) outperformed the other weighting schemes (Table \ref{table:DR}, rows 11-16).

\paragraph{Changing the model: stacked cylinders}
To challenge the generality of the results, a cylinder model was tested. This model consisted of three cylinders stacked along the longitudinal axis. Each cylinder had the same radius, but the cylinder lengths and scattering length densities varied (Figure \ref{fig:cyl}). This model was less symmetric than the core-shell model and represented a different contrast situation. However, the conclusion remained the same: the naive weighting scheme ($w_j =1$) provided the most accurate results, followed by the information-based weighting scheme ($w_j=N_{g,j}/M_j$), which were both much better than the reduced weighting scheme ($w_j=1/M_j$) (Table \ref{table:DR}). Notably, when fitting against simulated SAXS data with 2000 points and simulated SANS data with 50 points, only the naive weighting scheme ($w_j =1$) was superior to refinement against SAXS data alone. For the two other weighing schemes, the refined parameters became less accurate from inclusion of an additional SANS dataset with different contrast, but much fewer points (Table 1, bottom two rows). 

\begin{table}
	\caption{Average relative deviation of each weighting schemes for all condition described in the main text (lower deviation is better). Calculated as in equation \eqref{eqn:av_dev}. For the core-multishell model, the structural parameters $R_c$, $R_1$, $R_2$ and $R_3$ were included in the deviation metric, but nuisance parameters like scaling, background and contrasts were not. For the raspberry model, the core radius and the thickness of the first two layers were considered. For the stacked cylinder model, the structural parameters $R$, $L_1$, $L_2$ and $L_3$ were included in the deviation measure. Using bootstrapping, the 99\% confidence intervals were determined to be approximately 1\% across the different test cases, which is reflected in the number of significant digits displayed in the table. $M_N$ and $M_X$ are the number of points in the simulated SANS-like and SAXS-like datasets, respectively. *when the number of points in SAXS and SANS datasets are the same, then $w_j =1/M_j$ is equivalent to $w_j = 1$ .** The additional SANS contrast for the core contained 50 points.}
	\begin{tabular}{l l l l l l l }      
		\hline
		Core-multishell  model & $M_\mathrm{N}$:$M_\mathrm{X}$ & $w_j = 1$ & $w_j =1/M_j$ & $w_j=N_{g,j}/M_j$ & SANS & SAXS  \\
		\hline
		SAXS + SANS  & 300:300 & 4.8 & * & 4.8 & 8.1 & 20.4 \\
		SAXS + SANS  & 50:400 & 6.4 & 7.3 &  6.5 & 12.5 & 19.4 \\
		SAXS + SANS  & 50:900 & 6.2 & 7.8 & 6.6 & 12.7 & 21.5 \\
		SAXS + SANS  & 50:2000 &  4.7 & 7.1 & 5.8 & 12.6 & 14.3 \\
		add core contrast & 50:400** &  2.1 & 2.8 & 2.2 & 12.8 & 27.7\\
		add core contrast & 50:2000** & 1.2 & 2.7 & 1.8 & 12.8 & 15.3 \\
		add structure factor & 50:400 &  12.4 & 13.0 &  12.3 & 18.9 & 24.1 \\
		add structure factor & 50:2000 &  9.6 & 12.7 & 10.8 & 18.9 & 19.6 \\
		Raspberry-like surface & 50:400 & 50 & 55 & 52 & 77 & 65 \\
		Raspberry-like surface & 50:2000 & 45 & 52 & 50 & 69 & 60 \\
		SANS res. eff. ($\times1.0$) & 50:400 &  6.5 & 7.3 &  6.5 & 13.0 & 19.6 \\ 
		SANS res. eff. ($\times1.0$) & 50:2000 & 4.9 & 7.6 & 6.2 & 13.3 & 15.3 \\
		SANS res. eff. ($\times2.0$) & 50:400 &  6.8 & 7.8 &  7.0 & 13.3 & 19.4 \\
		SANS res. eff. ($\times2.0$) & 50:2000 & 5.1 & 8.9 & 7.2 & 13.8 & 15.5 \\
		SANS res. eff. ($\times3.0$) & 50:400 & 8.3 & 9.3 &  8.4 & 13.9 & 19.5 \\
        SANS res. eff. ($\times3.0$) & 50:2000 & 6.1 & 10.8 & 9.9 & 14.1 & 15.4 \\
		\hline
		Stacked cylinder model & $M_\mathrm{N}$:$M_\mathrm{X}$ & $w_j = 1$ & $w_j =1/M_j$ & $w_j=N_{g,j}/M_j$ & SANS & SAXS  \\
		\hline
		SAXS + SANS & 50:400 & 10.2 & 12.0 & 11.6 & 19.3 & 25.5 \\
		SAXS + SANS & 50:2000 & 5.1 & 10.4 & 9.8 & 18.0 & 8.0\\
		\hline
	\end{tabular}
	\label{table:DR}
\end{table}

\subsection{Effect of over- or underestimated errors}
To investigate the effect of poor error estimates, data were simulated again using the core-multishell model, but this time, the errors of either the SANS or the SAXS data were multiplied with a factor between 0.1 and 10, after they had been simulated. Thus, the reported errors of the simulated data did no longer reflect the fluctuations of the data around the true value. The errors ranged from highly underestimated (a factor of 0.1) to highly overestimated (a factor of 10). 

\begin{figure}
	\caption{Effect of over- or underestimated errors on parameter refinement. (A) Examples of simulated SANS data with over- or underestimated errors. (B-E) Radii of the core-multishell model when refined  50,000 times against SAXS and SANS data, with the latter having over- or underestimated the errors by a factor between 0.1 (highly underestimated) to 10 (highly overestimated).}
	\label{fig:errors}
	\includegraphics[width=0.96\linewidth]{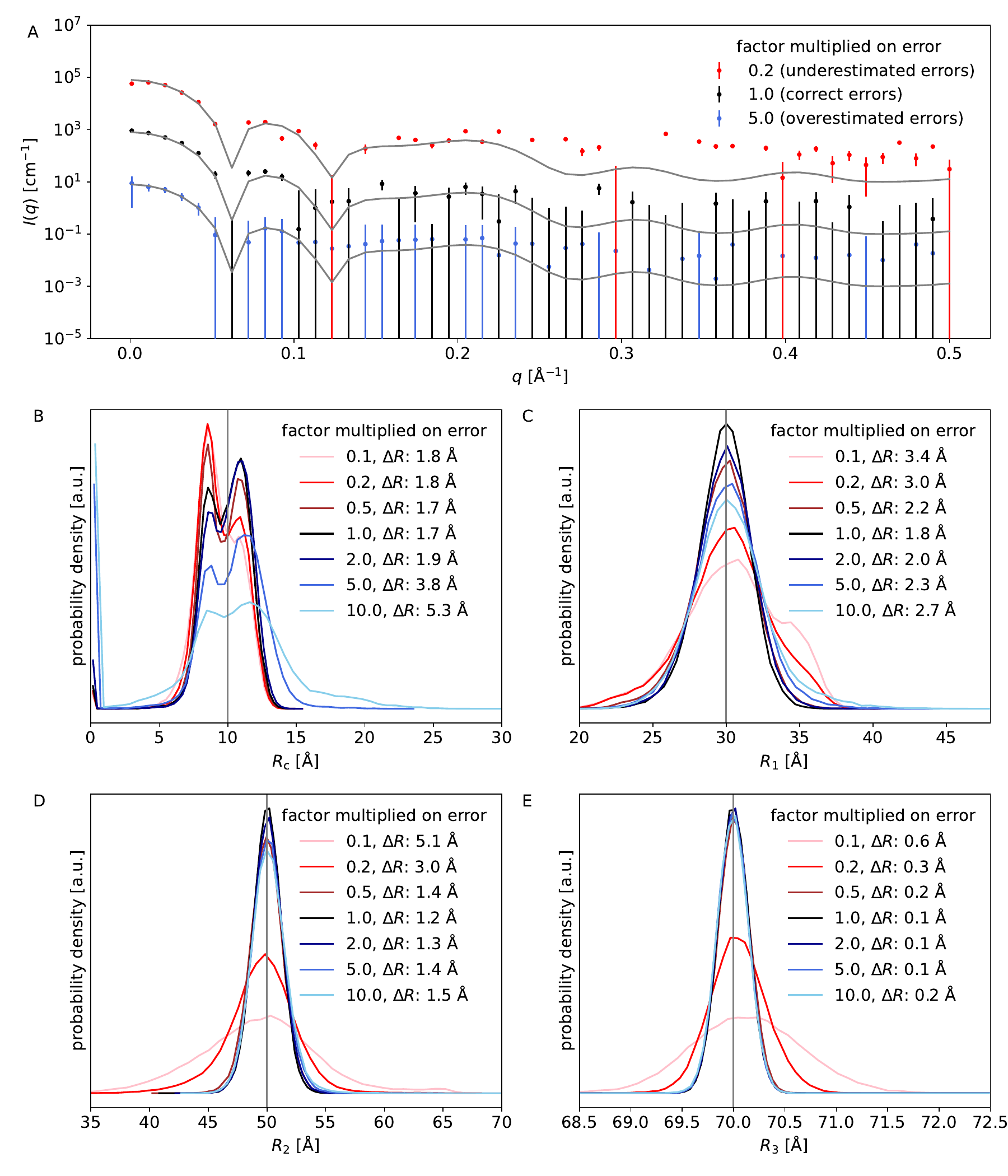}
\end{figure}

In the first round, the SAXS data were kept unchanged while the SANS errors were changed to be either underestimated or overestimated. The radii of the core-multishell model were then estimated against the SAXS and altered SANS data. Not surprisingly, the radii were determined most accurately when the errors were correct (Figure \ref{fig:errors}). 
Overestimation of SANS errors had severe effects on the core radius in the core-multishell model ($R_c$), because this parameter was predominantly determined by the SANS data. On the other hand, underestimation of the SANS errors had little on $R_c$, but made the estimation of the outermost radius, $R_2$ worse, as this  parameter was predominantly determined from the SAXS data, and too low SANS errors effectively gave to little weight to the SAXS data (Figure \ref{fig:errors}).
In the second iteration, the roles were shifted, and the errors in the SAXS data were varied, while keeping SANS errors at the correct level (Figure S6). In that case, the most severe effects were observed for $R_c$ when SAXS errors were underestimated. 
These results illustrate that over- or underestimation of errors can led to poorer estimates of the refined model parameters. The effect depends on the contrast situation, the signal-to-noise ratio of the datasets, and on the degree of over- or underestimation. Therefore, errors should be assessed and, if possible, corrected before model refinement against multiple SAXS/SANS contrasts \cite{Larsen2021-chi2,Smales2021}. 

\subsection{Inclusion of priors}
Now we turn our focus towards how prior information can be included in the modeling. In conventional model refinement, no prior distribution is explicitly attributed the parameters, but most fitting programs, allow the user to set a minimum and a maximum value for each parameter \cite{Kohlbrecher2022-SASfit,Ilavsky2009-Irena}. This is equivalent to applying a uniform distribution for each parameter. So far in this paper, we have used such uniform priors, only limiting the parameters to a certain range around the true value, and preventing negative values where relevant. The simplest alternative is Gaussian priors, which are defined by a mean $\mu_\mathrm{prior}$ and a standard deviation $\sigma_\mathrm{prior}$. Gaussian priors can be included using Bayesian refinement.  
It has previously been shown, that inclusion of Gaussian priors (as opposed to uniform priors) improves the robustness of the refinement \cite{Larsen2018-BayesFit}. However, this was only shown for the refinement against a single SAXS/SANS contrast. Multiple datasets can be fitted simultaneously by minimizing the sum:
\begin{align}
	\min\left[\left(\sum_{j=1}^{N_\mathrm{dataset}}w_j\chi^2_j\right) + \alpha S\right],
	\label{eqn:prior}
\end{align}
where $S$ represents the prior, and $\alpha$ is the effective weight given to the prior. 
To investigate the effect of the prior, the naive weighting scheme was used ($w_j=1$) on simulated data of core-multishell particles. The model parameters were co-refined against a SAXS-like dataset with 400 points and a SANS-like dataset with 50 points. 

\paragraph{Description of prior distributions.}
Three sets of Gaussian prior distributions were generated ("poor prior", "good prior" and "best prior"), where the best prior is the set of priors that are closest to the true values. The Gaussian priors were truncated, with the minimum and maximum values defined as being five standard deviations from the mean ($\mu\pm5\sigma$). For the radii, a lower limit of 0 was also set if $\mu-5\sigma<0$. A non-informative uniform prior was also generated for comparative analysis ($\mathrm{Uniform}_{5\sigma}$), which was constant between the upper and lower limits and zero outside this interval.

Prior values for the radii are given in Table \ref{table:R}. All priors had the same values for all other parameter, i.e. contrasts, scaling and background (Table \ref{table:p}).

\begin{figure}
	\caption{Radii of the core-multishell model were refined against SAXS and SANS data, using a non-informative uniform prior (red), a poor Gaussian prior (light blue), a good Gaussian prior (dark blue) or the best Gaussian prior (black). The probability distributions were normalized, such that their maximum value is unity.}
	\label{fig:prior}
	\includegraphics[width=0.96\linewidth]{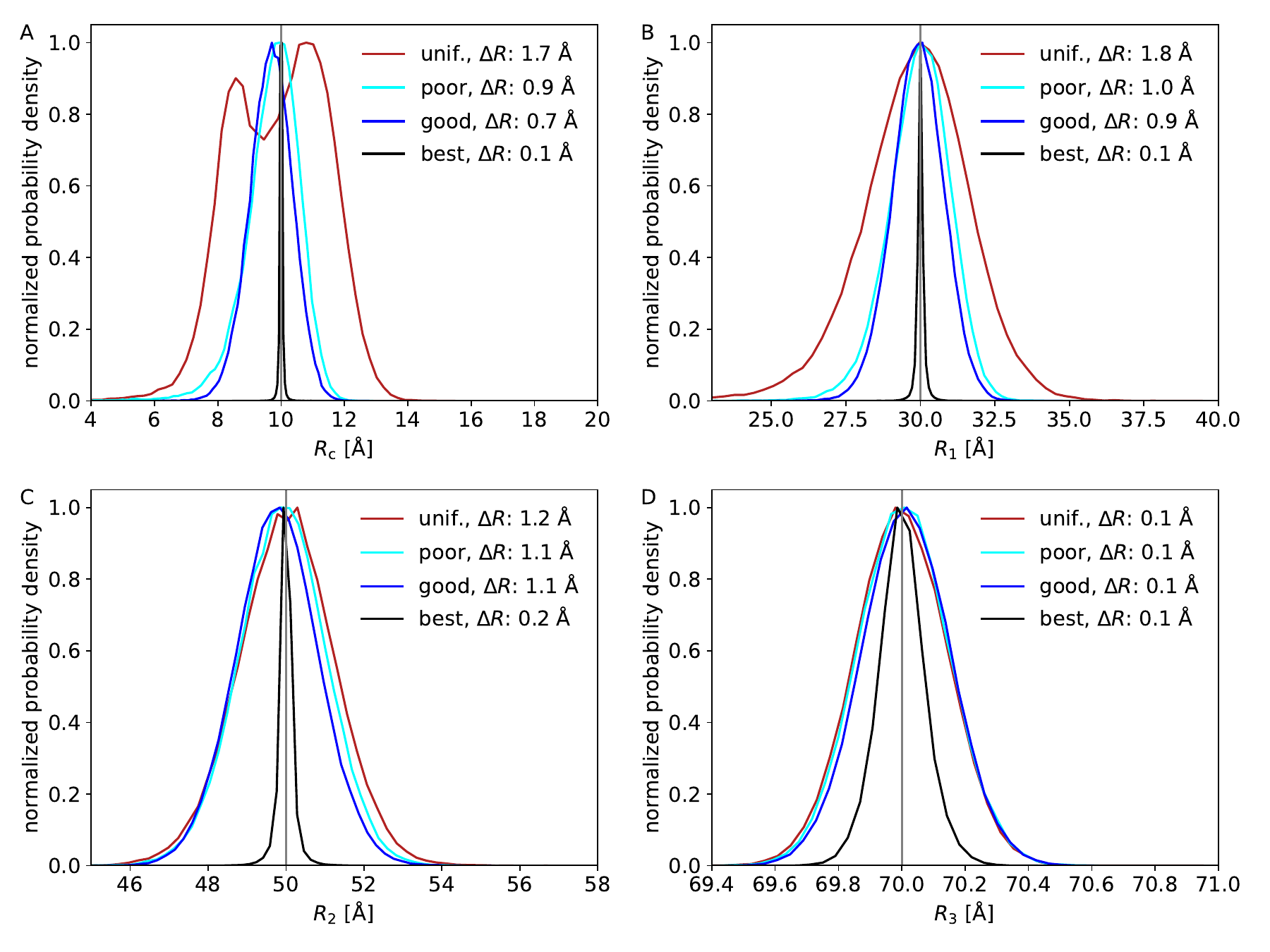}
\end{figure}

\paragraph{Gaussian priors improve the accuracy of the refined parameters.}
The estimates of $R_\mathrm{c}$, $R_1$ and $R_2$ were substantially improved by all tested Gaussian priors compared to the non-informative uniform prior (Figure \ref{fig:prior}). The best prior resulted in a very narrow distribution of refined values, although the prior width was relatively wide (Figure S7 and Table \ref{table:R}). The refinement of $R_3$, on the other hand, ws not improved by inclusion of Gaussian priors, as this parameter is very well defined by the data. Generally, the better a parameter was determined from data itself, the smaller was the effect of the prior. Importantly, the priors did not worsen the refined parameter values, even when the priors were relatively poor (Figure \ref{fig:prior}). 

\paragraph{Improving the uniform priors}
The uniform prior was stepwise improved by narrowing the upper and lower bounds from $\mu_\mathrm{best}\pm5\sigma_\mathrm{best}$ to  $\mu_\mathrm{best}\pm1/2\sigma_\mathrm{best}$. The results got increasingly more accurate, but even the narrowest uniform prior, gave substantially larger deviations than the best Gaussian prior (Figure S8). Remarkably, the poor prior resulted in smaller deviation compared to all uniform priors with minimum and maximum values of $\pm1\sigma_\mathrm{best}$ or higher.  This illustrates that Gaussian priors are better than uniform priors at guiding the minimization algorithm towards the correct solution, while being less restrictive.

\begin{table}
	\caption{True values and prior values for the radii of the core-multishell model. For the Gaussian priors, the mean ($\mu$) and standard deviation ($\sigma$) are given along with the upper and lower limits, which are $\mu\pm5\sigma$, or zero for the lower limit. For the uniform priors, the mean values, $\mu$, were used as the  initial value in the fit. For Uniform$_{5\sigma}$, the minimum and maximum values were the same as for the Gaussian priors, namely $\mu\pm5\sigma$. The other uniform priors are narrower, with subscript indicating the distance from $\mu$ to upper/lower limit.}
	\begin{tabular}{l l l l l}      
		Prior name    & $R_\mathrm{c}$  (min,max) [\AA]      & $R_1$  (min,max) [\AA]     & $R_2$  (min,max) [\AA] & $R_3$  (min,max) [\AA]   \\
		\hline
		True value & 10 & 30 & 50 & 70 \\
		Uniform$_{5\sigma}$      & $10$ (0,35)      & 30 (0,80)      & 50 (0,125) & 70 (0,170)     \\
		Uniform$_{4\sigma}$      & $10$ (0,30)      & 30 (0,70)      & 50 (0,110) & 70 (0,150)     \\
		Uniform$_{3\sigma}$      & $10$ (0,25)      & 30 (0,60)      & 50 (5,95) & 70 (10,130)     \\
		Uniform$_{2\sigma}$      & $10$ (0,20)      & 30 (10,50)      & 50 (20,80) & 70 (30,110)     \\
		Uniform$_{1\sigma}$      & $10$ (5,15)      & 30 (20,40)      & 50 (35,65) & 70 (50,90)     \\
		Uniform$_{\frac{1}{2}\sigma}$  & $10$ (7.5,12.5)  & 30 (25,35)     & 50 (40,55) & 70 (60,80)     \\
		Gaussian$_\mathrm{poor}$     & $5\pm5$ (0,30)     & $40\pm10$ (0,90)     & $45\pm15$ (0,120) & $90\pm20$ (0,190) \\
		Gaussian$_\mathrm{good}$     & $8\pm4$ (0,28)     & $35\pm10$ (0,85)    & $40\pm20$ (0,140) & $80\pm10$ (30,130)   \\
		Gaussian$_\mathrm{best}$      & $10\pm5$  (0,35)    & $30\pm10$ (0,80)     & $50\pm15$ (0,125) & $70\pm20$ 70 (0,170)   \\
	\end{tabular}
	\label{table:R}
\end{table}

\begin{table}
	\caption{True values and prior values for all model parameters except radii, which are given in Table \ref{table:R}). The same means ($\mu$) and standard deviations ($\sigma$) were used in all Gaussian priors. For the uniform priors, the means were used as initial guesses. In all priors, uniform and Gaussian, the upper and lower limits were $\mu\pm5\sigma$.}
	\begin{tabular}{l c c c}     
		Parameter    & True value  & $\mu\pm\sigma$ &  ($\mu-5\sigma,\mu+5\sigma$) \\
		\hline
		$(\frac{\Delta\rho_1}{\Delta\rho_\mathrm{c}})_\mathrm{SAXS}$ & 2 & $2.0\pm0.2$ & $(1.0,3.0)$  \\
		$(\frac{\Delta\rho_1}{\Delta\rho_\mathrm{c}})_\mathrm{SANS}$ & -0.1 & $-0.10\pm0.01$ & $(-0.15,-0.05)$ \\
		$(\frac{\Delta\rho_2}{\Delta\rho_\mathrm{c}})_\mathrm{SAXS}$ & 3 & $3\pm3$ & $(1.5,4.5)$ \\
		$(\frac{\Delta\rho_2}{\Delta\rho_\mathrm{c}})_\mathrm{SANS}$ & 0.1 & $0.10\pm0.01$ & $(0.05,0.15)$  \\
		$(\frac{\Delta\rho_3}{\Delta\rho_\mathrm{c}})_\mathrm{SAXS}$ & 4 & $4.0\pm0.4$ & $(2.0,6.0)$ \\
		$(\frac{\Delta\rho_3}{\Delta\rho_\mathrm{c}})_\mathrm{SANS}$ & 0.05 & $0.050\pm0.005$ & $(0.025,0.075)$  \\
		$a_\mathrm{SAXS}$  [cm$^{-1}$] & 0.5  & $0.50\pm0.05$  & $(0.25,0.75)$ \\
		$a_\mathrm{SANS}$  [cm$^{-1}$] & 0.8  & $0.80\pm0.08$  & $(0.1,0.9)$ \\
		$b_\mathrm{SAXS}$ [$10^{-4}$ cm$^{-1}$] & 0.1  & $0.1\pm100$  & $(-500,500)$ \\
		$b_\mathrm{SANS}$ [$10^{-4}$ cm$^{-1}$] & 1.0  & $1.0\pm100$  & $(-499,501)$ \\		
	\end{tabular}
	\label{table:p}
\end{table}

\section{Experimental example: cardian clock protein complex}
In an elegant study by Yunoki et al., the structure of the circadian clock protein complex was determined with SAXS and SANS \cite{Yunoki2022}. The hexameric protein complex consists of multiple domains, KaiA, KaiB, and KaiC, and in the SANS experiment, the KaiB and KaiC domains were matched out. SAXS and SANS data were thus complementary and could exclude different structural candidates, in particular the SANS data excluded two proposed structure classes (Type 2 and Type 3) \cite{Yunoki2022}. The data were deposited on SASBDB with IDs SASDNK2 (SANS data) SASDNJ2 (SAXS data). 

Here, the data were used to showcase the use of priors and weights in simultaneous fitting of multiple SAS datasets. First, the experimental errors were assessed using the BIFT algorithm \cite{Larsen2021-chi2}. The SANS errors were assessed to be correct, whereas the SAXS errors were assessed to be slightly underestimated, so these were rescaled by a factor of 1.6, to obtain a better balance between SAXS and SANS data. 

\begin{figure}
	\caption{(A) Representative snapshots from the two simulated ensembles, with the CHARMM36-IDP force field leading to asymmetric structures, and with the AMBER14SB force field, leading to symmetric structures. The central part of the protein complex was matched out in SANS (KaiB and KaiC) \cite{Yunoki2022}. (B) Simultaneous fitting of SAXS (with rescaled errors, SASDNJ2, red) and SANS data (SASDNK2, blue), displaying also the prior, with equal amounts of the two structural ensembles. Normalized residuals displayed below the fits.}
	\label{fig:MD}
	\includegraphics[width=0.95\linewidth]{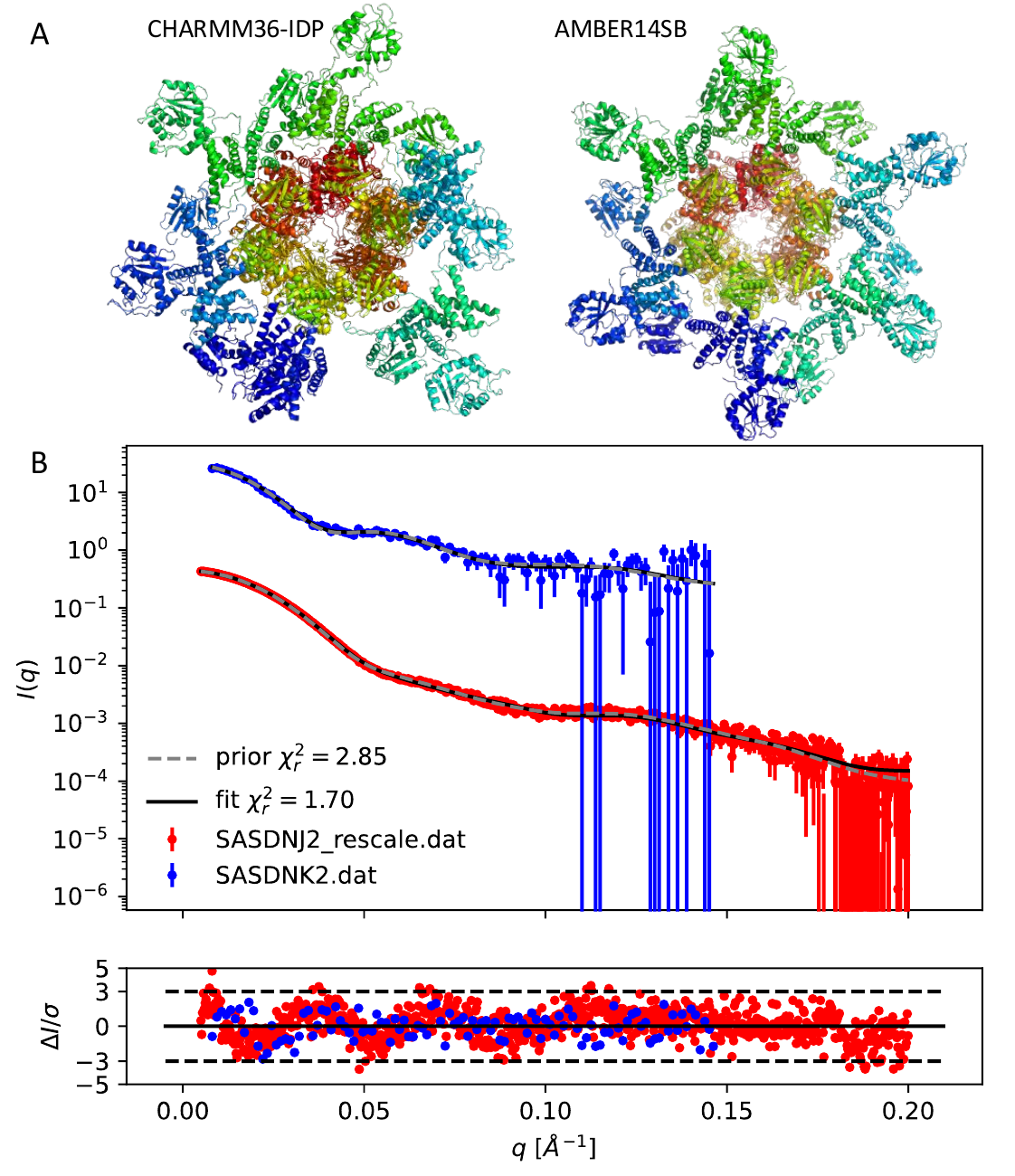}
\end{figure}

A model structure deposited at the SASBDB entry was used as initial structure and a 100 ns simulation was run with two different force fields to probe various structural arrangements and their consistency with the SAXS and SANS data. The first force field, AMBER14SB provides an ensemble of relatively symmetric structures, whereas the second force field, CHARMM36-IDP was developed for intrinsically disordered proteins and breaks the symmetry of the complex (Figure \ref{fig:MD}).  The symmetric structural ensemble generated with the AMBER14SB force field was consistent with the data, with a reduced $\chi^2$ value of 1.7 for the simultaneous fit. The asymmetric structural ensemble generated with the CHARMM36-IDP force field was less consistent with the data ($\chi^2$ 6.6). However, there could be a minor fraction of asymmetric structures in the sample, as observed for other protein multimers \cite{Johansen2022}. To determine whether this was the case for the circadian clock protein complex, a mixture of the structural ensemble was used to fit the data, where $f_\mathrm{sym}$  and $f_\mathrm{asym}$ are the fraction of structures from the symmetric (AMBER14SB) or asymmetric (CHARMM36-IDP) ensembles, and the calculated scattering from each ensemble are $ I_\mathrm{sym}$ and $I_\mathrm{asym}$, respectively. The mixed scattering can then be described as:
\begin{align}
	I_\mathrm{model}(q) = s \left(\frac{f_\mathrm{sym}}{f_\mathrm{asym}} I_\mathrm{sym}(q) + I_\mathrm{asym}(q)\right)
\end{align}
where $s$ is an overall scaling parameter. A non-informative log-normal prior distribution was used for the stoichiometric ratio, $\log(f_\mathrm{sym}/f_\mathrm{asym}) = 0\pm2$, corresponding to assuming that half of the ensemble structures are symmetric and half are asymmetric. 

By simultaneous fitting of the SAXS and SANS data using the naive weight scheme ($w_j =1$), the stoichiometry was refined to 90\% [88, 91] (68\% confidence interval) symmetric structures from the AMBER force field ensemble  and 10\% [9, 12] asymmetric structures from the CHARMM36-IDP force field ensemble, with a $\chi^2_r$ of 1.7 for the total simultaneous fit. $\chi^2_r$ was 1.8 for the simultaneous fit against the SAXS data and it was $\chi^2_r$ 1.2 for the fit to SANS. 

Using the reduced weight scheme ($w_j =1/M_j$), the stoichiometry was instead refined to 89\% [15,98] symmetric and 11\% [2, 85] asymmetric with the same goodness of fit as above, but much higher uncertainty on the refined parameters. 

Refining against SAXS data alone gave the same result as for the naive weighting scheme, whereas refinement against SANS alone gave 77\% [57, 89] symmetric and 23\% [11, 43] asymmetric structures.

If the SAXS errors were not rescaled, the resulting stoichiometry (and confidence interval) were, in this case, essentially unchanged, but with a larger $\chi^2_r$ of 4.1 for the fit (and 4.5 for SAXS and 1.2 for SANS).

That is, in this example, SAXS is dominating in discriminating between the two structural ensembles. But this was not obvious, and using optimal weighting ensures that the most accurate solution is robustly found. The structural conclusion is that the addition of the asymmetric structure does not improve the fit to data compared to only using the symmetric ensemble, which supports the modeling strategy taken by Yonoki et al. \cite{Yunoki2022}, namely using the AMBER14SB force field. 

\section{Discussion}

\subsection{Why is the model refinement not dominated by the dataset with many datapoints?}
Even when one dataset had 2000 datapoints and the other only 50 datapoints, the refined parameters were still affected by both datasets. This is because the data contained orthogonal information. For some structural domains the contrast was low in SAXS and high in SANS. Therefore, an additional dataset can contain much structural information albeit having a low signal-to-noise ratio. On the other hand, if the contrast situation is similar in multiple small-angle scattering datasets that are simultaneously fitted, then the refined parameters will be dominated by the dataset having the better signal-to-noise ratio \cite{Pedersen2014-info,Larsen2020-tia1,Larsen2021-chi2}.

When datapoints are statistically independent, no additional weighting is necessary, i.e., the naive weighting scheme ($w_j=1$) leads to the most accurate result, as demonstrated with the simulated data. Oversampling of data, i.e. that the number of datapoints exceeds the number of Shannon channels, which was the case for the simulated data, does not lead to statistical dependency. However, it is crucial to avoid operations in the data reduction process that introduce dependence, or take these into account in the error propagation \cite{Heybrock2023}. 

\subsection{When experimental errors are ill-defined}
Error estimates are important for getting the correct balance between multiple datasets, when doing model co-refinement (Figure \ref{fig:errors}). Methods have previously been presented to identify, and in some cases, correct over- or underestimated errors \cite{Larsen2021-chi2,Smales2021}. However, there is only limited work on how to identify systematic errors, e.g., from non-optimal buffer subtraction  \cite{Shevchuk2017}. This becomes particularly important when high flux, long exposure times, and stable samples at high concentrations allow the statistical errors to reach a level, where the fluctuations in data are dominated by errors that are not accounted for. Such effects may likely be the cause that the SAXS dataset used in the experimental example (SASBDB ID: SASDNJ2) was assessed to have underestimated errors by the BIFT algorithm. Goodness of fit measures that exploit runs tests do not depend on statistical errors and are therefore valuable tools for identifying variations that are not reflected in the counting statistics-based errors \cite{Franke2015,Kofinger2021}.

\section{Conclusion}
The most optimal weighting scheme for simultaneous fitting of multiple datasets is simply $w_j=1$. That is, the sum of the (non-reduced) $\chi^2$ values should be minimized. This was compared to a weighting scheme with the information content taken into account ($w_j=N_{\mathrm{g},j}/M_j$) and with a weighing scheme relying on reduced $\chi^2$ values rather than  $\chi^2$ values ($w_j=1/M_j$). The naive weighting scheme ($w_j=1$) gave most accurate results, in particular when there was substantial difference in the number of points in each included dataset.

Inclusion of Gaussian priors gave more accurate refinement of structural parameters than uniform priors. This has previously been demonstrated for single SAXS datasets \cite{Larsen2018-BayesFit}, but here it was demonstrated that this was also the case when simultaneously fitting multiple SAXS or SANS datasets.

Implementing optimal strategies for data analysis, as proposed in this study, is a pragmatic approach to enhance accuracy of structural refinements. These strategies require minimal resources compared to the immense work that is needed to prepare samples and built and maintain SAXS and SANS instruments. Still, they offer substantial improvement in the accuracy of the refined parameters, and ultimately aid scientists reaching more accurate and consistent conclusions.         

\section{Acknowledgments}
The author thanks Wojtek Potrzebowski for insightful comments on the manuscript. The project was funded by the Carlsberg Foundation (grant  CF19-0288) and the Lundbeck Foundation (grant R347-2020-2339). 

\bibliographystyle{iucr}
\bibliography{ref}

\end{document}